\documentclass[acus]{pac99}
\usepackage{epsfig}
\newcommand{\bit}{\begin{Itemize}}
\newcommand{\eit}{\end{Itemize}}
\begin{document}

\thispagestyle{empty}

\onecolumn

\begin{flushright}
{\large
SLAC--PUB--8232\\
August 1999\\}
\end{flushright}

\vspace{.8cm}

\begin{center}

{\LARGE\bf
Dynamic Aperture Studies for SPEAR~3\footnote
{\normalsize{Work supported by
Department of Energy contract  DE--AC03--76SF00515.}}}

\vspace{1cm}

\large{
J.~Corbett,
Y.~Nosochkov,
J.~Safranek \\
Stanford Linear Accelerator Center, Stanford University,
Stanford, CA  94309\\

\medskip

A.~Garren \\
UCLA Center for Advanced Accelerators, Los Angeles, CA 90095 \\
}

\end{center}

\vfill

\begin{center}
{\LARGE\bf
Abstract }
\end{center}

\begin{quote}
\large{
The SSRL is investigating an
accelerator upgrade project to replace the present 130 nm$\cdot$rad
FODO lattice with an 18 nm$\cdot$rad double bend achromat lattice:
SPEAR~3. In this paper, we review the methods used to maximize
the SPEAR~3 dynamic aperture including optimization of linear optics,
betatron tune, chromaticity and coupling correction, and
effects of machine errors and insertion devices.
}
\end{quote}

\vfill

\begin{center}
\large{
{\it Presented at the 1999 IEEE Particle Accelerator Conference (PAC99)\\
New York City, New York, March 29 -- April 2, 1999} \\
}
\end{center}

\newpage

\pagenumbering{arabic}
\pagestyle{plain}

\twocolumn

\title{
DYNAMIC APERTURE STUDIES FOR SPEAR~3\thanks
{Work supported by the Department of Energy Contract
DE--AC03--76SF00515 and the Office of Basic Energy Sciences, Division
of Chemical Sciences.}
}

\author{
J.~Corbett,
\underbar{Y.~Nosochkov\thanks{E-mail: yuri@slac.stanford.edu.}},
J.~Safranek, SLAC, Stanford, CA 94309 \\
A.~Garren, UCLA, Los Angeles, CA 90095
}

\maketitle

\begin{abstract}
The SSRL is investigating an 
accelerator upgrade project to replace the present 130 nm$\cdot$rad
FODO lattice with an 18 nm$\cdot$rad double bend achromat lattice: 
SPEAR~3. In this paper, we review the methods used to maximize 
the SPEAR~3 dynamic aperture including optimization of linear optics, 
betatron tune, chromaticity and coupling correction, and 
effects of machine errors and insertion devices.
\end{abstract}
\vspace{-2mm}
\section{INTRODUCTION}
SPEAR~3 is a machine upgrade project under study at SSRL \cite{y:upgrade}. 
It aims at replacing the current 130 nm$\cdot$rad FODO lattice with an
18 nm$\cdot$rad double bend achromat (DBA) lattice. To
reduce the cost of the project and to use the existing synchrotron light
beam lines, the new design \cite{y:cdr} closely follows the racetrack 
configuration of the SPEAR tunnel, with the magnet positions fit to the 
18 magnet girders. The 3 GeV lattice has two-fold symmetry and periodicity 
with two identical arcs and two long straight sections. 

\begin{figure}[tb]
\includegraphics{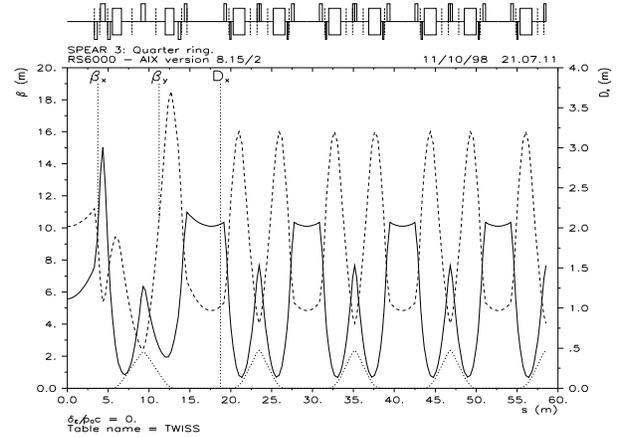}
\vspace{58mm}
\caption{Optics of one quadrant of SPEAR~3.}
\label{y:f:qring}
\vspace{-4mm}
\end{figure}

The lattice functions for one quarter of the ring are shown in 
Fig.~\ref{y:f:qring}. Similar to other light source rings, it has been 
found advantageous to use combined function bends to relax 
the optics and reduce sextupole strength.

Though the DBA design has an advantage of a high brightness 
beam, the strong focusing increases beam sensitivity to machine
errors and generates larger chromaticity. The stronger
sextupoles increase the amplitude dependent and non-linear chromatic 
aberrations and reduce the dynamic aperture. It is especially important to 
maximize the horizontal size of dynamic aperture to minimize
the Touschek effect and allow large injection oscillations.

In the following sections we review the lattice optimization
and tracking studies. The tracking simulation was done using LEGO 
\cite{y:lego}. The dynamic aperture was calculated at
the symmetry point between arc cells where 
$\beta_{x}/\beta_{y}=10.1/4.8$~m. 

\vspace{-2mm}
\section{ERROR FREE LATTICE}
The DBA cell was made to fit the existing 11.7 m cell length with 
the magnet positions constrained by the existing photon beam lines and
$\sim$3 m space for the insertion devices (ID). This results in a compact DBA 
design with stronger focusing and increased sensitivity to machine 
errors. Vertical focusing added to the bends increased the separation of $x$
and $y$ focusing and reduced strength of the quadrupoles and 
sextupoles. Further reduction of the sextupole strengths was achieved by 
moving the bends as far apart as possible.

The phase advance per arc cell was chosen to be near 
$\mu_{x}$=0.75$\times$2$\pi$ and $\mu_{y}$=0.25$\times$2$\pi$. 
This provides conditions for local cancellation of: 
1) geometric aberrations from arc sextupoles located $-I$ apart, and 
2) first order chromatic beta waves from sextupoles and quadrupoles 
located $\pi$/2 apart, as well as systematic quadrupole errors.
With this choice, the total tune would be near $\nu_{x}$=14.5 
and $\nu_{y}$=5.5. To move the working tune away from the 1/2
resonance and to minimize resistive wall impedance effects,
the tune was moved into the lower quarter on the tune plane 
($\nu$$<$1/2). 

The matching cell optics was optimized by performing a horizontal 
dynamic aperture scan across the matching cell phase advance. 
The dynamic aperture including $\delta$=$\frac{\Delta p}{p}$ up to $\pm$3\% 
was maximized at $\mu_{x,m}=0.78$$\times$2$\pi$ and 
$\mu_{y,m}=0.42$$\times$2$\pi$ per matching cell.

To minimize the effect of low order betatron resonances the 
working tune was chosen slightly below 1/4, away from the 3rd and 
4th order resonance lines. The final 
choice ($\nu_{x}=14.19$, $\nu_{y}=5.23$) was based on favorable 
horizontal injection conditions and the results of dynamic aperture tune 
scan. With the chosen tune, the phase advance per arc cell is 
$\mu_{x}=0.7907$$\times$2$\pi$ and $\mu_{y}=0.2536$$\times$2$\pi$.

As mentioned previously, the chosen phase advance in the arc cells 
provides conditions for local compensation of chromatic and geometric 
aberrations. This scheme would work optimally for the number of arc cells 
of 4$\times${\it{integer}}. With only 7 cells per arc, constrained by the
SPEAR geometry, the correction is not complete. 

The study showed that chromaticity correction with only 2 sextupole 
families did not provide adequate dynamic aperture for
particles up to $\delta$=$\pm$3\%. Since the 2 families only compensate 
linear chromaticity, the off-momentum aperture is limited by the non-linear 
chromatic effects. A significant amount of non-linear chromaticity is
generated in the matching cells which break periodicity of the 14 arc cells
and contribute $\sim$20\% to the total chromaticity. 
Two additional sextupole families (SFI, SDI) placed in the matching cells 
reduced the non-linear terms by a factor of 3 and significantly improved 
the off-momentum aperture.

The matching cell sextupoles also generate geometric aberrations and 
therefore have to be kept relatively weak in order to preserve the 
on-momentum aperture. The optimum strengths of the SFI, SDI
were evaluated through a horizontal aperture scan 
versus SFI, SDI strengths.

To increase optical separation between the SF and SD sextupoles
two other options were studied. In one option, the SD 
was combined with part of the adjacent bend. This increased $\beta_y$ 
but reduced dispersion at the SD which led to smaller dynamic aperture. 
In the second study, the SF was combined with the center quadrupole QFC. 
This increased dispersion and $\beta_x$ at the SF and reduced its
strength, but dynamic aperture did not improve.
\vspace{-2mm}
\section{MACHINE ERRORS}
In tracking simulations, we included random main field errors, random 
and systematic multipole errors, and random alignment errors. In addition, 
a skew octupole component was added to skew quadrupoles combined with
sextupoles, and a feed-down multipole field was included due to a
large trajectory in the bends \cite{y:feed}.

The alignment rms errors for bends, quads and sextupoles used in 
the study were: $\Delta x$,$\Delta y=200$ $\mu$m, roll $=500$ $\mu$rad.
The rms main field errors due to differences in magnetic core length 
were assumed to be (1-2)$\cdot 10^{-3}$.

The multipole field errors were defined in terms of ratio of the 
multipole field $\Delta B_{n}$ (normal or skew) to the main magnet field 
$B$ at radius $r$, where $n=1,2,\ldots$ is the multipole order starting 
with a bend. The normal systematic and random rms values $\Delta B_{n}/B$ 
used in the study are listed in Tables~1,2. Conservatively
large values were specified for $n=3,6,10,14$ multipoles on the
quads.

\begin{table}[tb]
\begin{center}
\caption{Systematic rms multipole field errors.}
\medskip
\begin{tabular}{lccc}
\hline
\textbf{Magnet} & \boldmath{$r(mm)$} 
                & \boldmath{$n$} 
                & \boldmath{$\Delta B_{n}/B$} \\
\hline
Dipole & 30 & 2 & $1\times 10^{-4}$ \\
       & & 3-14 & $5\times 10^{-4}$ \\
Quadrupole & 32 & 6,10,14 & $5\times 10^{-4}$ \\
Sextupole & 32 & 4 & $-8.8\times 10^{-4}$ \\
          & & 5 & $-6.6\times 10^{-4}$ \\
          & & 9 & $-1.6\times 10^{-3}$ \\
          & & 15 & $-4.5\times 10^{-4}$ \\
\hline
\end{tabular}
\label{y:t:sysmult}
\end{center}
\vspace{-8mm}
\end{table}
\begin{table}[tb]
\begin{center}
\caption{Random rms multipole field errors.}
\medskip
\begin{tabular}{lccc}
\hline
\textbf{Magnet} & \boldmath{$r(mm)$}
                & \boldmath{$n$} 
                & \boldmath{$\Delta B_{n}/B$} \\
\hline
Dipole & 30 & 2 & $1\times 10^{-4}$ \\
Quadrupole & 32 & 3,6,10,14 & $5\times 10^{-4}$ \\
           &    & 4,5,7-9,11-13 & $1\times 10^{-4}$ \\
Sextupole & 32 & 5 & $1.5\times 10^{-3}$ \\
          & & 7 & $4.8\times 10^{-4}$ \\
\hline
\end{tabular}
\label{y:t:ranmult}
\end{center}
\vspace{-8mm}
\end{table}
The skew quadrupoles physically combined with chromatic sextupoles 
provided an efficient coupling correction.
In total, we used 24 skew quads arranged in 4 families.
The induced vertical dispersion was small and far outweighed by the 
improved aperture. As a future option, the vertical dispersion can be 
corrected as well by using more independent skew quads.
The combined skew quad and sextupole gives rise to a skew
octupole field that was systematically added in the simulations.
\vspace{-2mm}
\section{DYNAMIC APERTURE WITH ERRORS}
For tracking simulations, LEGO first generates a set of magnet errors,
then applies tune, orbit, chromaticity and coupling corrections,
and finally tracks the particles. For tune correction, 
two families of doublet quads in the cells were used. 
Linear chromaticity typically was adjusted to zero with the SF, SD
sextupoles. An RF voltage of 3.2 MV was used to generate 
synchrotron oscillations.

\begin{figure}[b]
\vspace{-20mm}
\includegraphics{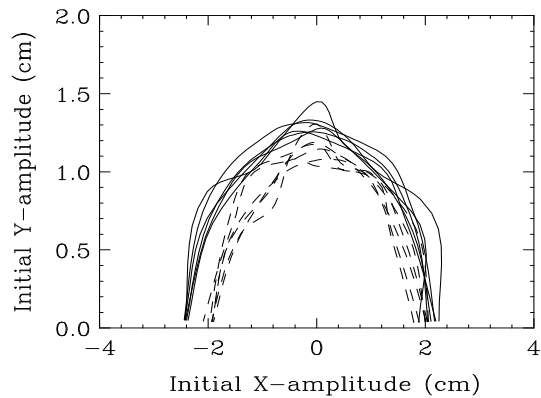}
\vspace{68mm}
\caption{Dynamic aperture for 6 seeds of machine errors
for $\delta=0$ (solid) and 3\% (dash).}
\label{y:f:seed}
\end{figure}

The resultant dynamic aperture without ID's for 6 random seeds of machine 
errors is shown in Fig.~\ref{y:f:seed}. The horizontal dynamic aperture 
is 18-20 mm for $\delta$=$\pm$3\% momentum range. 
This provides sufficient aperture for a long Touschek lifetime and
injection oscillations.

In real machines, the linear chromaticity $\xi$=$\Delta\nu/\delta$ is 
typically set slightly positive. Since non-zero chromaticity increases
the momentum dependent tune spread, the off-momentum particles would 
cross more betatron resonances during synchrotron oscillations. 
Resonance effects can reduce the $\delta$$\neq$0 aperture if $\xi$ is large. 
The $\delta$=0 aperture is affected only by the increased sextupole strength.

\begin{figure}[t]
\includegraphics{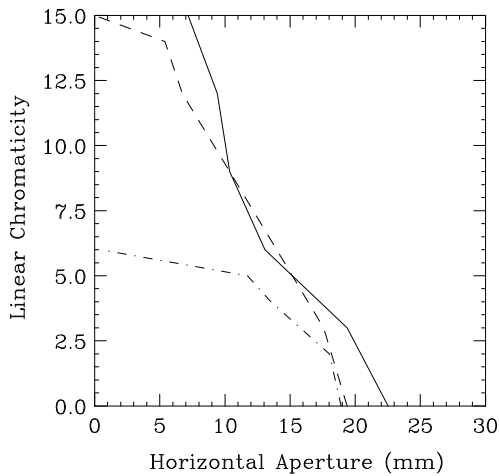}
\vspace{63mm}
\caption{Horizontal dynamic aperture versus linear chromaticity for
$\delta=0$ (solid), 1\% (dash) and 3\% (dot-dash).}
\label{y:f:axchrom}
\vspace{-4mm}
\end{figure}

Fig.~\ref{y:f:axchrom} shows dependence of horizontal dynamic aperture on
$\xi$ for various $\delta$ (the vertical dependence is 
similar). In this study, the chromaticity was set equal in both planes. 
Clearly, for the SPEAR~3 tune the particles lose stability near 1/2
resonance, when $\Delta\nu$$\approx$-0.2.
Though the dynamic aperture for the core beam (small $\delta$) is not 
much affected, the Touschek lifetime can be reduced for
$\xi$$>$5.

In the study, typical $\beta$ distortions after correction were 
$\Delta\beta/\beta$$<$$\pm$10\%. In a real machine larger 
modulations can occur. To verify the effect of large $\beta$ modulation 
the quad field errors in two matching quad 
families were increased to generate
$\Delta\beta_{x}/\beta_{x}$$\approx$$\pm$30\% and
$\Delta\beta_{y}/\beta_{y}$$\approx$$\pm$20\%. 
The observed aperture reduction was about 15\%. Though this 
aperture is still adequate to operate the machine, such large 
errors have to be resolved in practice.

The effect of large orbit distortions was studied by including an
additional set of uncorrected rms alignment errors.
For rms orbit distortions of $\Delta x$=3 mm, $\Delta y$=1.5 mm
a maximum of 2 mm reduction of dynamic aperture was observed for
the particles within $\delta$$<$$\pm$3\%.

Similarly, large sextupole misalignments were studied while the orbit
was well corrected. In simulation, 1 mm rms sextupole misalignments were
included which are equivalent to about 10 times the focusing
errors from ring quadrupoles. Of the 6 seeds studied, 5 cases showed 
$>$17 mm horizontal dynamic aperture for $\delta$=0 particles. 
The vertical aperture was larger than the $\pm$6 mm ID chamber size. 
At $\delta$=3\%, the horizontal aperture remained $>$13 mm.

\begin{figure}[t]
\includegraphics{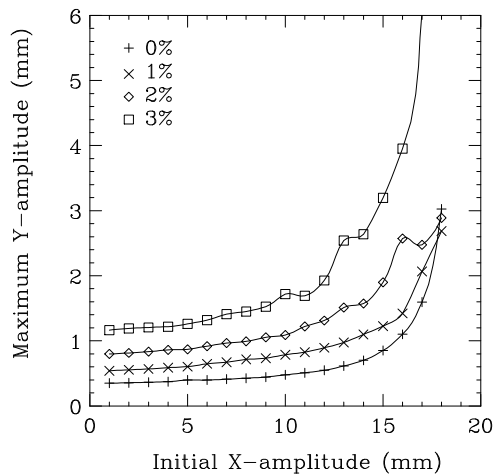}
\vspace{63mm}
\caption{Peak vertical excursion vs initial horizontal
amplitude for $\delta=0,1,2,3\%$.}
\label{y:f:xycoupl}
\vspace{-4mm}
\end{figure}

The actual physical aperture can be a limiting factor for
a beam lifetime. In the vertical plane, SPEAR has two ID's
with $y$=$\pm$6 mm vacuum chamber. In case of strong coupling, the large 
off-momentum horizontal motion can be transferred into vertical amplitude 
which could increase beam loss at the vertical physical aperture.
To study this effect, we monitored the maximum vertical excursion at 
an ID location as a function of initial horizontal amplitude. 
Fig.~\ref{y:f:xycoupl} shows the peak 
$y$-orbit averaged over 6 random sets of machine errors with 
$\delta$=0-3\% energy oscillations. At 10 mm injection oscillations, 
the induced vertical amplitude is below 2 mm and should not limit the 
beam lifetime.

In addition to chromatic effects, sextupoles generate geometric 
aberrations such as amplitude dependent tune shift and high order 
resonances. Clearly, the dynamic aperture
would reach its maximum if the geometric aberrations
were zero. One way to verify this limit is to track particles
with fixed $\delta$=0 and without sextupoles. The
simulation showed that on-momentum aperture could be 40\% larger
without sextupole aberrations. 

The geometric effects can be 
reduced by using 'harmonic' sextupoles placed in non-dispersive regions. 
Based on analysis in \cite{y:hsex} we tested a scheme of two harmonic 
sextupole families placed in the arc cells. The sextupole strengths
were scanned to maximize dynamic aperture.
The harmonic correction reduced the amplitude dependent tune shift by 
about 40\% and the error free dynamic aperture improved by 10-15\%. 
With machine errors, however, the improvement reduces 
to a minimum. Currently, the harmonic correction is not included in 
the design.

The fields in insertion devices further reduce the dynamic
aperture. The detailed report of wiggler effects in SPEAR~3 is presented 
in \cite{y:wig}. In summary, the first order wiggler focusing will
be locally compensated using cell quadrupoles on either side of a wiggler.
Simulations showed that with corrected wigglers, included systematic
multipole errors and intrinsic wiggler fields up to dodecapole, the
dynamic aperture reduces to about 18 mm and 8-9 mm in $x$ and $y$ 
planes, respectively.
\vspace{-2mm}
\section{CONCLUSIONS}
Tracking studies combined with optimization of SPEAR~3 lattice show  
sufficient dynamic aperture for 10 mm injection oscillations and 
$>$100 hrs of Touschek lifetime.
The dynamic aperture results have also been confirmed
by M. Borland with the tracking code ELEGANT.
\end{document}